\documentclass[sigconf,nonacm]{acmart}

\AtBeginDocument{%
  \providecommand\BibTeX{{%
    \normalfont B\kern-0.5em{\scshape i\kern-0.25em b}\kern-0.8em\TeX}}}

\usepackage{pgf}
\usepackage{tikz}
\usetikzlibrary{arrows,automata}
\usetikzlibrary{positioning}
\usepackage{hyperref}
\begin{document}

\title{Quantum Shuttle: Traffic Navigation with Quantum Computing}

\author{Sheir Yarkoni}
\email{sheir.yarkoni@volkswagen.de}
\affiliation{%
  \institution{Volkswagen Data:Lab}
  \streetaddress{Ungererstraße 69}
  \city{Munich}
  \country{Germany}
}
\additionalaffiliation{%
  \institution{Leiden Institute for Advanced Computer Science, Leiden University}
  \streetaddress{Niels Bohrweg 1}
  \city{Leiden}
  \country{The Netherlands}
  \postcode{2333 CA}
}

\author{Florian Neukart}
\affiliation{%
  \institution{Volkswagen Advanced Technologies}
  \streetaddress{Main Street}
  \city{San Francisco}
  \state{California}
}
\additionalaffiliation{%
  \institution{Leiden Institute for Advanced Computer Science, Leiden University}
  \streetaddress{Niels Bohrweg 1}
  \city{Leiden}
  \country{The Netherlands}
  \postcode{2333 CA}
}

\author{Eliane Moreno Gomez Tagle}
\affiliation{%
  \institution{Volkswagen Data:Lab}
  \streetaddress{Ungererstraße 69}
  \city{Munich}
  \country{Germany}
}

 \author{Nicole Magiera}
 \affiliation{%
   \institution{Volkswagen Data:Lab}
   \streetaddress{Ungererstraße 69}
   \city{Munich}
   \country{Germany}
 }

\author{Bharat Mehta}
\affiliation{%
  \institution{Hexad GmbH}
  \streetaddress{Stralauer Allee 8}
  \city{Berlin}
  \country{Germany}}

\author{Kunal Hire}
\affiliation{%
  \institution{Hexad GmbH}
  \streetaddress{Stralauer Allee 8}
  \city{Berlin}
  \country{Germany}}

\author{Swapnil Narkhede}
\affiliation{%
  \institution{Hexad GmbH}
  \streetaddress{Stralauer Allee 8}
  \city{Berlin}
  \country{Germany}}
  
  
\author{Martin Hofmann}
\affiliation{
  \institution{Volkswagen AG}
  \streetaddress{Main Straße}
  \city{Wolfsburg}
  \country{Germany}
}

\renewcommand{\shortauthors}{Yarkoni, et al.}

\begin{abstract}
The Web Summit conference in Lisbon, Portugal, is one of the biggest technology conferences in Europe, attended by tens of thousands of people every year. The high influx of people into Lisbon causes significant stress on the city's transit services for the duration of the conference. For the Web Summit 2019, Volkswagen AG partnered with the city of Lisbon for a pilot project to provide quantum computing-based traffic optimization. A two-phase solution was implemented: the first phase used data science techniques to analyze the movement of people from previous conferences to build temporary new bus routes throughout the city. The second phase used a custom Android navigation app installed in the buses operated by Carris, powered by a quantum optimization service provided by Volkswagen that connected to live traffic data and a D-Wave quantum processing unit to optimize the buses' routes in real-time. To our knowledge, this is the first commercial application that depends on a quantum processor to perform a critical live task. 
\end{abstract}



\maketitle

\section{Introduction}
As cities around the world continue to grow in both size and population, traffic congestion becomes an increasingly prevalent problem~\cite{EUmobility}. This is especially apparent during events that congregate large numbers of people for specific periods of time. For example, conferences, sporting events, and festivals can cause temporary but significant disruption to the cities' transportation systems, resulting in delays for the residents of those cities~\cite{trafficmanagementmajorevents, citytraffic}. A key issue is that permanent transportation infrastructure, such as rail lines or roads, are costly and slow to modify given the temporary nature of these events. In light of this, the advent of smart traffic management systems offers possible improvements to existing transportation systems with minimal overhead in regards to implementation. Some requirements for such systems include the management of the mobility flows in real- or near to real-time using flexible and modular software components. At the same time detailed traffic forecasts should also be considered as well as the identification of the best scenarios to manage congestion, road closures, disruptions such as events or accidents, and accurately simulate the interaction of all vehicles, mobility modes and pedestrians~\cite{componentsofcongestion}. Using the data collected and the insights obtained from the predictions, cities in the urban mobility world of the future will be able to steer the demand of the whole transportation network. Building such a traffic management system that leverages large amounts of data in real time poses a significant challenge using contemporary technologies, necessitating the use of emerging technologies to overcome these challenges. One of the most promising candidates of such a technology is quantum computing, which has been gaining popularity among both academic and industrial circles in hopes of solving complex optimization problems in a time-relevant manner~\cite{ttt, quantumsupremacy, jspQA}. Companies such as Google, IBM, D-Wave Systems, and others, have started offering access to quantum processing units (QPUs) for experimentation in hopes of accelerating application development for the new technology. However, despite these impressive advancements in hardware, real-world applications have yet to be developed outside of small proof-of-concept demonstrations. 


Of these demonstrations, a small number have shown how to translate various traffic-related optimization problems to be suitable for quantum computers~\cite{tfopaper, densotfo}. Specifically, it was shown how to minimize intersections between candidate routes for vehicles by formulating the task as a quadratic unconstrained binary optimization (QUBO) problem. The QUBO problem was then solved using a QPU provided by D-Wave Systems, which uses quantum annealing to perform  heuristic optimization~\cite{nishimori}. In this paper we describe a connected traffic management system that was operated during the Web Summit 2019 conference in Lisbon, Portugal, inspired by the optimization problem formulated in~\cite{tfopaper}. 
We specifically address and solve two main problems:\\

\begin{enumerate}
    \item How do we design customized bus routes to avoid traffic congestion during big events?
    \item How would one build a real-time production application using a quantum computer to manage such a traffic system? \\
    
\end{enumerate}

Our application, named the ``Quantum Shuttle'', was live for a period of four days, navigating nine buses operated by the transit authorities of Lisbon (Carris) between the Web Summit conference center and the city center of Lisbon from November 4-7, 2019. The buses were navigated by a custom Android application which was dependent upon the D-Wave QPU performing the route optimization task. The buses followed the routes as assigned by the QPU in the form of turn-by-turn instructions. We consider this work the first commercial application of quantum computing, where a live service was offered to the general public powered at its core by a quantum processor.

The rest of the paper is structured as follows: Section~\ref{section:mobility} explains how the bus routes were extracted from movement data, Section~\ref{section:qws} details how our cloud computing-based navigation service was designed and implemented, and Section~\ref{section:testing} describes the testing phases and the lessons learned from them. Sections~\ref{section:websummit2019} and~\ref{section:conclusions} present the results of the live run and our conclusions, respectively.

\section{Inferring bus routes from movement data}
\label{section:mobility}

In order to understand how to improve the overall traffic conditions during the Web Summit dates in Lisbon, a comprehensive movement stream analysis was performed in collaboration with the PTV Group\footnote{\href{PTV}{https://www.ptvgroup.com/en/}}, a traffic and logistics consulting company. The main goal of the analysis was to provide recommendations to the public transport operator of Lisbon, Carris, upon fleet allocation, capacity, and scheduling, to reinforce their existing network during the Web Summit conference in the form of a new mobility service-- the Quantum Shuttle service-- operated by Carris during the days of the conference. The movement stream analysis was divided into two main phases: the first phase, commissioned to PTV Group, included the setup of the transport model of Lisbon, using the PTV Visum software. This was based on a four-step transport modelling approach: trip generation, trip distribution, modal split, and traffic assignment. The second phase was performed by the Volkswagen Data:Lab mobility team: deriving bus stop locations, bus routes, and schedules for the Quantum Shuttle based on the given demand for mobility in the transport model from PTV.

\begin{figure}[h]
  \centering
  \includegraphics[width=\linewidth]{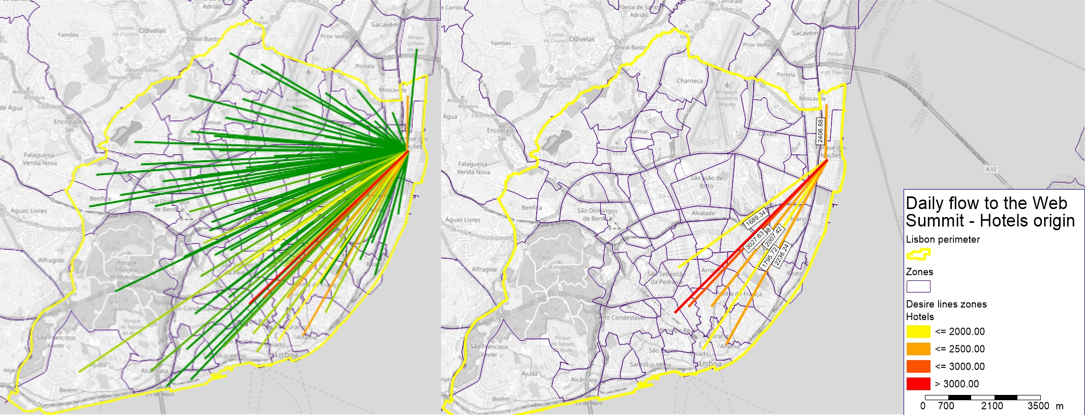}
    \caption{Left: Origin-destination matrices from the Web Summit conference to the city center. Right: Selected OD matrices for hotel- and room rental-related trips. Visualization provided by PTV Visum software.}
  \Description{Origin-destination pairs of movement data (left) with hotel PoI trips (right).}
  \label{figure:odmatrices}
\end{figure}

The goal of the first phase was to obtain origin/destination (OD) matrices, detailing the volume of movement streams from the Web Summit conference venue (Altice arena) to different zones of the city of Lisbon on an hourly basis. The results from the PTV analysis showed a total of 225 OD matrices in a study area of 93 zones throughout Lisbon, shown in Figure~\ref{figure:odmatrices}. The data which served as an input into the traffic model created by PTV included demographic census data, mobility behavior from surveys, Lisbon traffic counts, floating car data, mode of choice and network models from the city, for both dates within and outside the Web Summit time frame.

To better understand the nature of the traffic conditions, movement data was segmented according to the trip purpose, such as: family, airport, hotels, and others. Data was segmented by hour of day, using the peak day of visitors for Web Summit (Wednesday) as a reference. The OD matrices also differentiated the mode of choice, either travel by public transport or private transport. The second phase of the movement analysis derived the exact coordinates of new or existing bus stops to be included in the new Quantum Shuttle service and reinforce the existing public transport network. For this, locations of the biggest traffic jams caused
by conference traffic were analyzed. By comparing the transport model
of days within and outside of the conference time frame, movement streams leading into those traffic jams were identified as the source, and traced back to their zones of origin. Only movement streams with more than 2000 trips per day were used (Figure~\ref{figure:odmatrices}, right) were included in this analysis. All trip purposes other than family trips and homes were included in the OD matrices, to yield the likeliest representation of potential users of the Quantum Shuttle service. As a result, 12 high demand zones towards and from the Web Summit were identified. Trips per hour were extracted for each of the high demand zones, separated into public transport and private transport modes. Of the 12 original high demand zones, only zones with more than 250 trips per hour were selected as candidates for covering the target passenger base by the Quantum Shuttle fleet. Four zones were finally selected to cover the morning and evening demand. 

\begin{figure}[h]
  \centering
  \includegraphics[width=0.9\linewidth]{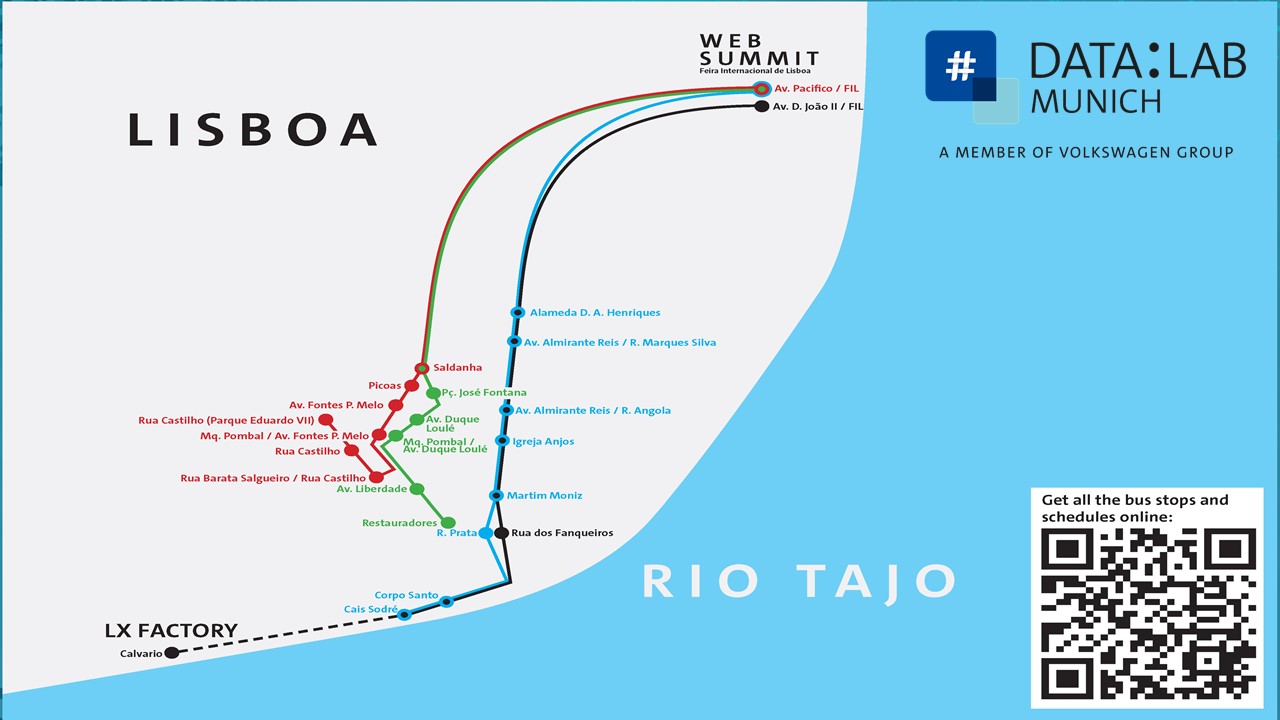}
    \caption{Final selection of bus lines for the Web Summit Quantum Shuttle service.}
  \Description{Quantum Shuttle bus lines.}
  \label{figure:shuttleroutes}
\end{figure}

For each of the zones, a geospatial analysis correlating the points-of-interest (such as hotels and private room rentals) with the existing public transport network in Lisbon was conducted and discussed with Carris. The results suggested that the pick-up and drop-off bus stops needed to be no more than 2-5 minutes walking distance for the service to be attractive to customers. To accomplish this, three express bus lines were proposed to serve the demand from the selected zones to the Web Summit: a red, green, and blue line, with a total of 23 bus stops. One express line was dedicated for the return traffic from the conference venue to the city center (the black line). A schematic of the map used on display at the Web Summit conference is shown in Figure~\ref{figure:shuttleroutes}. The red and green bus lines operated only during the morning period, whereas the black and blue lines operated both in the morning and evening. Bus departures were scheduled every 30 minutes for all lines. For the morning, two lines (red and green) covered the demand in the northern part of the city center, picking up visitors along 7 dedicated bus stops and meeting at the farthest point of the line at the Saldahna roundabout. From this point to Web Summit, the visitors were no longer picked up and the bus was navigated solely using the quantum navigation service. For the black and blue lines, the portion of the routes between the Web Summit and Alameda station were navigated using the quantum navigation service. 

To determine the schedules of the fleet, we identified the peak traffic demand in the selected zones was from 09:00-10:00 and 10:00–11:00, with 6314 trips towards Web Summit. For the evening demand (the return trips to the city center), peak hours were from 16:00-17:00 and 17:00-18:00 with 7600 trips leaving the Web Summit. The results also highlighted that the zone with the lowest public transport usage was zone 75 (Santa Maria Maior-Castelo), with $45\%$ of total trips being public transit. This indicated that the modal split in this zone can be heavily improved relative to other zones with higher average public transit usage ($65\%$ and above). Given the performed analysis for each of the four selected zones, and considering the estimated demand in both the morning and evening, a total of 9 buses was proposed to Carris as a recommendation for the Quantum Shuttle fleet volume. 



\section{Building the quantum web service}
\label{section:qws}
Building a functioning web service that uses a cloud-based quantum processor while providing meaningful navigation optimization imposes a specific set of both conditions and constraints. By setting the goal of navigating buses in real-time, we require that a live connection between three different services be consolidated simultaneously-- the Android navigation app, the traffic data, and the quantum optimization. We briefly describe the content and scope of each component, then explain how they were combined in the final web service. For brevity, from here on we refer to the cloud-based part of the Quantum Shuttle service as the Quantum Web Service, or QWS.

\subsection{Bus navigation Android app}
\label{section:hexad}
Development of the Android app was divided into two parts: the front-end (visualization for the bus drivers) and the back-end (server-side communication to the QWS). We explain separately how the two parts were implemented, then consolidated, to satisfy the demands of the Quantum Shuttle navigation. 

\textbf{Front-end.} The front-end of the bus navigation comprised of an Android application, visualizing the turn-by-turn navigation, operated on an Android tablet. The main role of the app was to plot the custom routes provided by the QWS and initiate turn-by-turn navigation with voice instructions. Additional functionality was built into the app to allow bus drivers to start/stop the current and next trips they were meant to follow, as well as to track the current location of the buses in the fleet to allow for the optimization of the route selection. Trips were also ended automatically once a vehicle was within 15 meters of its defined destination. The visualization portion of the app was built using the Mapbox SDK\footnote{\href{Mapbox}{https://www.mapbox.com/}} to render the custom routes obtained by the QWS. 

In order to make the app useful for live navigation, multiple issues had to be overcome in its implementation. Most importantly, the current location of the moving buses needed to be reconciled with the custom routes received from the QWS. Thus, if the vehicle had already passed the assumed starting point returned by the QWS, the app would form U-turns and loops in order to catch the missed point, making the app unusable in a live setting. To solve this, the app used a projected future location for the vehicle instead of the live location when broadcasting to the route optimization. However, it was necessary to ensure the future location was a valid GPS point along the route. To do this, an algorithm was developed using the Mapbox SDK to decide the future location considering properties such as geographic line, vehicle movement, bearing angles, etc. While this solution allowed us to project the future location accurately, the live location of the vehicle was still required by the QWS for monitoring and data collection purposes. Therefore, both locations, live and projected, were stored locally on the Android devices, and sent to the back-end component periodically (explained in the next section). 

In the event that device connectivity was lost during the trip, the active trip was manually ended (purposefully or accidentally), or the QWS experienced errors, it was vital for the Android app to maintain valid turn-by-turn directions for the bus to reach its destination. To accomplish this, a fallback mechanism was implemented: a static route was loaded into every device for all the trips (which was overwritten by the custom routing of the QWS) to be used in the event of no live route optimization, and a cache was used to maintain the route provided by the QWS in the event of lost connection. Thus, even if a bus became isolated, it could still complete its trip. 




\textbf{Back-end.} The back-end of the navigation app acted as an interface between the Android app and the QWS. The back-end was used to send the collected data of all the vehicles of the Quantum Shuttle fleet at fixed time intervals to QWS over HTTP. Responses from the QWS were also distributed by the back-end to their respective vehicles.

The most important role of the back-end was to keep the data flow in sync between the vehicles and the QWS, as both synchronous and asynchronous communication protocols were used. The QWS was designed in such a way that it accepted a single consolidated request consisting of the location information for all the vehicles, both live and projected. It was the role of the back-end software to consolidate all the locations it received from all the vehicle devices at different frequencies, and send it to the QWS. This meant that the back-end maintained an index of each vehicle's request and response, to ensure data integrity. Because of the different requirements of the live location tracking and the projected locations, the requests submitted by the back-end to the QWS were separated between two destinations on the QWS: /update and /optimize. The /update request submitted the live location of each Quantum Shuttle vehicle to the QWS on a 30 second interval, while the projected location (and in return the customized route returned) was submitted to the /optimize URL at an interval of 120 seconds. The necessity of splitting the requests to separate URLs and their respective timing was discovered during the testing phase, explained in Section~\ref{section:testing}. 

Lastly, the back-end was also responsible for determining the difference between two subsequent optimized routes for the same vehicle. The optimized route was sent to the vehicles' app only when the optimized route was changed. Therefore, every time a route was changed, a matching of the new coordinates was performed, and redundant GPS coordinates were removed (for example, new sets of coordinates describing the same route as before). This step was critical in the performance and user experience perspective: adjusting the routes on the Android devices would delay instructions and cause ambiguity for the bus drivers. Additionally, performing these adjustments on the back-end reduced the risk of the front-end crashing due to unforeseen errors. 

\subsection{Traffic and route optimization}
\label{section:tfo}
The approach used for custom navigation of the vehicles followed an approach based on~\cite{tfopaper}. At the start of every trip and at regular time intervals until completion of each trip, multiple candidate routes needed to be generated between the current location of each bus in the system to its assigned destination. These routes also needed to be traffic-aware to reflect the current conditions of the city road network. To accomplish this, we used a live traffic services provider, HERE Technologies~\footnote{\href{HERE}{https://www.here.com/}}. Using their routing API~\footnote{\href{HERE routing}{https://developer.here.com/}}, we were able to generate between 3-5 traffic-aware candidate routes per vehicle at every optimization step with minimal overhead. 

It is important to note that since the vehicles were operated in parallel in different parts of Lisbon, different routes were likely to be suggested for each vehicle in the system at every optimization step. Often, however, subsets of these suggested routes overlapped, necessitating the optimization of the routes' selection to minimize congestion. In this scenario, identical GPS points were returned from the HERE API describing the shape of the overlapping portion of the routes. Therefore, the GPS points were used directly to form the optimization problem, instead of the road segments as in~\cite{tfopaper}. The QUBO formulation of the objective function is then:

\begin{equation}
    \mathrm{Obj} = \sum_{p \in P} \mathrm{cost}(p) + \lambda \sum_i \left( \sum_j q_{ij} - 1 \right)^2,
\end{equation}

where $q_{ij}$ are the binary decision variables associated with vehicle $i$ taking route $j$, $P$ is the set of all GPS points that overlap in the suggested routes, $\lambda$ is a scaling factor ensuring only one route is selected per vehicle in the QUBO minimum, and $\mathrm{cost}(p)$ is the cost function associated with each GPS point $p$ in $P$:
\begin{equation}
    \mathrm{cost}(p) = \sum_{q_{ij} \in B(p)} \left( q_{ij} \right)^2.
\end{equation}

Here, $q_{ij}$ is as before, and $B(p)$ is the set of all binary variables that contain the GPS point $p$. Thus, the final selection of routes in the optimum of the QUBO represent the routes that minimally overlap with all other selected routes. 


\subsection{D-Wave QPU access}
\label{section:qpu}
Quadratic unconstrained binary optimization (QUBO) problems are defined on binary variables $\{0, 1\}$, and are equivalent to minimizing Ising Hamiltonians, which use spin variables $\{-1, 1\}$. This has been shown to be NP-hard in the worst case~\cite{barahona}, and such problems are often solved using heuristic algorithms~\cite{qubo}. Cutting edge quantum computing architectures, such as gate model and annealing QPUs, can represent Ising Hamiltonians natively, making  quantum algorithms an attractive tool for heuristic optimization of hard problems. There are many well-known combinatorial optimization problems that have been reduced to QUBO/Ising form~\cite{lucas}. 

For live traffic navigation, our quantum optimization algorithm must meet specific conditions. Because of the time-sensitive nature of traffic navigation, our solution needed to respond with valid solutions to the optimization problem quickly. The algorithm also needed to handle varying sizes and complexity of the traffic flow optimization (TFO) problem, as it needed to optimize the route selections of the vehicles automatically at regular intervals. Due to these real-world conditions, we used the D-Wave 2000Q QPU and its respective software stack to deploy our solution. The physics motivating the quantum annealing algorithm used by D-Wave QPUs is beyond the scope of this work, and extensive background can be found in literature~\cite{manufacturedspins, entanglement}. We deployed three different methods of using quantum annealing to solve the traffic flow optimization problem formulated in Section~\ref{section:tfo}. We briefly explain each method, how it was implemented, and evaluate them based on our navigation application.

\textbf{Direct embedding.} The most straightforward approach to solving QUBOs with a D-Wave quantum annealer is by minor embedding the graph directly to the topology of the QPU. The latest D-Wave QPU, the D-Wave 2000Q, has a limited-connectivity hardware graph, called Chimera. In order to solve arbitrarily-structured QUBO problems, multiple physical qubits can be chained together using additional constraints to represent logical variables. This technique is well-studied in literature~\cite{choi} and the D-Wave software tools have the ability to do this built-in. The benefit of this approach is speed-- even with the overhead of transforming the TFO problem to a QPU-compatible graph, using the QPU at the fastest annealing time ($1\mu s$ to obtain a single sample) still returned valid solutions to the problem. This process can be performed on the order of tens or hundreds of milliseconds. However, there are two drawbacks to this method. Firstly, by employing extra physical qubits to describe the problem (and by exclusively using the QPU directly), we restrict the size of problems that can be solved to a maximum of 2000 qubits. This restriction would prevent us from fully automating our system for large numbers of vehicles or routes. Secondly, qubits are analog devices, and therefore have a limit on the precision with which a problem can be described before data becomes indistinguishable from background noise. This means that the quality of the solutions obtained by the QPU degrade as the problem size increases, again preventing us from obtaining useful solutions to the TFO problem with many variables~\cite{mis}. During the development phase of the project we tested various configurations of the direct embedding approach, and found that this method was suitable for up to 10 cars with 5 routes each. 

\textbf{Hybrid algorithms.} One method of circumventing the issues of direct embedding is to build a hybrid quantum-classical algorithm. Hybrid algorithms are implemented on classical computers, and use a QPU as part of an inner loop to perform a specific task that improves the algorithm's ability to converge to optimal solutions~\cite{qbsolv, denso}. The advantage of this approach is that the use of the QPU in the algorithm can be tailored to the QPU's strength, making better use of a limited resource. However, run-time is sacrificed in waiting for the QPU's response to continue the iterative classical procedure. The work in~\cite{tfopaper} employed a hybrid algorithm that used a 64 variable fully-connected graph as the inner loop for the QPU to optimize sub-problems~\cite{qbsolv}. However, as mentioned previously, minor embedding dense problems significantly degrades the QPUs performance, which in this case still came at the cost of waiting for the QPU responses. In light of this, we developed a custom hybrid algorithm to make better use of the QPU in a timely manner. Our algorithm performed the same tabu search in the outer classical loop as in~\cite{qbsolv}, but instead of using a single 64 variable sub-problem, we found sub-graphs within the TFO problem that were already Chimera structured, thus circumventing the embedding issue. We were able to deploy a hybrid algorithm similar to~\cite{denso} but without employing chains in the sub-problem. Our method allowed us to increase the throughput of sub-problems to the QPU. However, due to time constraints, we were not able to parallelize the implementation so that it could continually run independent of the request/response portion of the QWS. Therefore, the hybrid algorithm was restarted every time a new route optimization was requested. This incurred significant overhead time, delaying the response to an unacceptable level for live use.

\textbf{Hybrid Solver Service.} As part of D-Wave's online cloud service, in addition to direct QPU access, a state-of-the-art hybrid algorithm is also being offered, currently in alpha testing. This service, named the Hybrid Solver Service (HSS), is tailored to solve large, arbitrarily structured QUBOs with up to 10,000 variables. The disadvantage of this method is that we cannot control the exact method with which the QPU is used, instead we use the HSS as an optimization black-box. Access to the HSS was provided through the same API as the D-Wave QPU, which allowed us to integrate it in to the QWS in a modular way. By offloading the overhead associated with starting the hybrid algorithm to the D-Wave remote server, we were able to reduce the response time significantly compared to the other approaches. That the HSS can solve problems significantly larger than the QPU also allows us to seamlessly scale up our QWS to handle hundreds of vehicles with tens of route options for future applications. The HSS provided the best compromise of speed, accessibility, and performance, and was used for the live Quantum Shuttle navigation during the Web Summit conference.

\subsection{Putting everything together}
\label{section:aws}
We briefly present an overview of the navigation service as a whole, detailing the various components comprising of the system and how they communicate, for review and clarity. A diagram of the system is presented in Figure~\ref{figure:qws}. 
The components are color-coded according to their roles and developers: blue components were developed by Hexad GmbH, the gray components by the Volkswagen Data:Lab, and the white components provided by external parties. \\

\usetikzlibrary{arrows, positioning}

\tikzset{
    qws/.style={
           rectangle,
           rounded corners,
           draw=black, very thick,
           minimum height=18em,
           minimum width=11em,
           inner sep=2pt,
           text centered,
           label={[label distance=-10mm]:#1}
           },
    external/.style={
          rectangle,
          rounded corners,
          draw=black, very thick,
          inner sep=3pt,
          text centered,
          },
  app/.style={
          rectangle,
          rounded corners,
          draw=black, very thick,
          inner sep=2pt,
          text centered,
         },
  num/.style={
          circle,
          draw=black, very thick,
          fill=grey,
          inner sep=1pt,
          text centered,
         },}
\pgfdeclarepatternformonly{soft horizontal lines}{\pgfpointorigin}{\pgfqpoint{100pt}{1pt}}{\pgfqpoint{100pt}{3pt}}%
{
  \pgfsetstrokeopacity{0.3}
  \pgfsetlinewidth{0.1pt}
  \pgfpathmoveto{\pgfqpoint{0pt}{0.5pt}}
  \pgfpathlineto{\pgfqpoint{100pt}{0.5pt}}
  \pgfusepath{stroke}
}

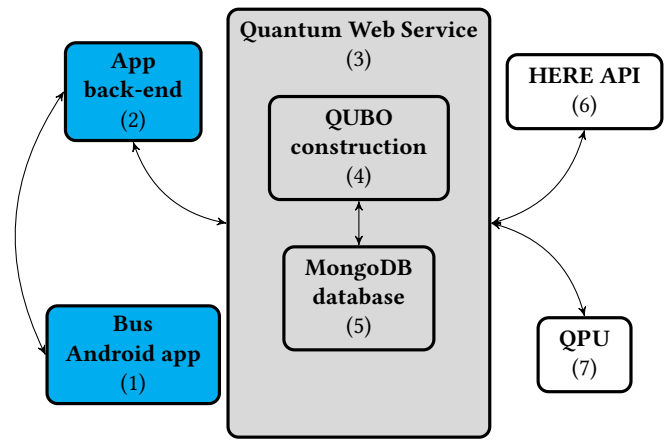
\begin{figure}[h]
  \centering

\begin{tikzpicture}[>=stealth']

 \node[fill=gray!30, qws={
 \begin{tabular}{c}
 \textbf{Quantum Web Service}\\(3)
 \end{tabular}}] (QWS) at (0, 0) {
 };
 
 \node[external] (HERE) at (3, +1.75) {
    \begin{tabular}{c}
        \textbf{HERE API} \\ (6)
    \end{tabular}
 };
 \node[external] (QPU) at (3, -1.75) {
    \begin{tabular}{c}
     \textbf{QPU} \\ (7)    
    \end{tabular}
 };
 \node[external] (QUBO) at (0, 1) {
    \begin{tabular}{c}
       \textbf{QUBO} \\
       \textbf{construction} \\ 
       (4)
    \end{tabular}
 };

 \node[external] (mongo) at (0, -1) {
    \begin{tabular}{c}
       \textbf{MongoDB} \\
      \textbf{database} \\ 
       (5)
    \end{tabular}
 };

 \node[fill=cyan, app] (bus) at (-3, -1.75) {
 \begin{tabular}{c}
        \textbf{Bus} \\
        \textbf{Android app} \\
        (1)
 \end{tabular}
 };
 
 \node[fill=cyan, app] (hexad) at (-3, +1.75) {
    \begin{tabular}{c}
        \textbf{App} \\
        \textbf{back-end} \\ (2)
    \end{tabular}
 };
  
\path [<->] (hexad.south) edge [bend right] (QWS.west);

\path [<->] (bus.west) edge [bend left] (hexad.west);

\path [<->] (QUBO.south) edge (mongo.north);

\path [<->] (QWS.east) edge [bend right] (HERE.south);

\path [<->] (QWS.east) edge [bend left] (QPU.north);


 
\end{tikzpicture}

  \caption{Diagram detailing the QWS and its interactions with the other components in the Quantum Shuttle navigation system, as per the text. \label{figure:qws}}
\end{figure}

\textbf{Blue components.} 
\begin{itemize}
    \item Component~(1), the front-end Android tablet application, provided visual turn-by-turn navigation with vocal instructions to the Quantum Shuttle bus fleet and their drivers. The bus locations were sent every 30 seconds to Component~(2). 
    \item Component~(2) was the Android app back-end, and submitted the POST requests to the QWS, Component~(3). Component~(2) consolidated the locations of the vehicles and submitted them to~(3) via the /update URL every 30 seconds; this component also formed the POST request to the Component~(3) /optimize URL for the route optimization, interpreted the response, and sent the new routes back to~(1).
\end{itemize}

\textbf{Gray components.} 
\begin{itemize}
    \item Component~(3) is the framework hosted on AWS (exposed using Flask\footnote{Flask is a minimalist Python framework for making web-apps, and can be found at: \href{Flask}{https://www.palletsprojects.com/p/flask/}.}) and served as the central consolidation point of the other components. In the event of an /update POST request from~(2), the QWS updated Component~(5), the MongoDB database used to store the data. In the event of an /optimize POST request from Component~(2), Component~(6) was accessed to request the suggest routes, and that data was passed to Component~(4) to construct the traffic flow optimization QUBO problem, then stored in the database using~(5). The QUBO was submitted to the D-Wave~HSS, Component~(7). Component~(3) then interpreted the results of the optimization, stored them via~(5), and pushed the selected routes back to~(2).
    \item Component~(4) is the Python module that implemented the traffic flow optimization QUBO formulation described in Section~\ref{section:tfo}.
    \item Component~(5) is the Python module wrapped around the MongoDB database; accessing and writing data from the QWS to the database was performed by this component.
\end{itemize}

\textbf{White components.} 

\begin{itemize}
    \item Component~(6) is the HERE Technologies traffic/routing API.
    \item Component~(7) is the D-Wave HSS (or other QPU-based services), accessible via HTTPS.
\end{itemize}

\section{Testing phase}
\label{section:testing}
In preparation for the Web Summit, multiple trial runs were performed, in both Wolfsburg, Germany (Volkswagen AG headquarters) and Lisbon, Portugal. In this section we describe the two different test scenarios and the lessons learned from each. 

\subsection{Initial tests: Wolfsburg}
The first testing phase occurred in Wolfsburg in August 2019, where a small number of cars (1-3) were driven between various origin/destination pairs using the Android app. The goal of this testing phase was to ensure the proper integration of the Android navigation app, quantum API, and the MongoDB database used to keep track of active trips and record results. During testing, several key observations were made, which were used to modify the system.

{\bf Location tracking.} As mentioned in Section~\ref{section:qws}, it was necessary to both track the live positions of the vehicles, as well as specify a new ``origin'' per vehicle for every optimization request. In order to ensure that the locations used for optimization didn't result in infeasible or unrealistic route selections, a projected location was used for each vehicle. Specifically, given a route in the form of a sequence of GPS coordinates, a GPS point further along along the route was passed as the origin for the next optimization step. During the initial testing phase, this position was also used to track the locations of the vehicles. However, this proved to be problematic, since the update frequency was faster than the change in the projected location, making it impossible to track the locations of the vehicles. To solve this, the live location and the location used to determine the new routes were separated: location updates were sent to a different URL independently from the optimization requests every 30~seconds, with the live locations now stored server-side after every update.

{\bf Optimization interval.} Before testing, the route optimization occurred every 60 seconds. During testing it was observed that this frequency was too fast, resulting in different routes being assigned to the vehicles every time an optimization problem was solved. Furthermore it was observed that, given a new route A after the optimization, a \emph{different new route} B would sometimes be suggested while navigating to route A. This would have made navigating buses with real passengers impractical. It was discovered that this phenomenon occurred due to two main reasons: firstly, quantum annealing is a heuristic optimization algorithm, meaning a QPU implementing such an algorithm is not guaranteed to return the same solution every time it is queried. In the event that multiple minimal solutions to the optimization problem have equal cost, the QPU may return any of the solutions~\cite{temperature}. Secondly, because of the frequent optimization requests, the vehicles' projected positions often stayed the same between optimization requests, causing the same optimization problem to be formulated in successive requests. After testing various optimization intervals (both faster and slower), 120~seconds between optimization requests resolved the issue, allowing for sufficient time to change the projected locations of the vehicles. 

\subsection{Final tests: Lisbon}
\label{section:finaltests}
The second testing phase occurred during September 2019, with a small number of Carris buses and drivers in Lisbon. The final bus routes selected as per Section~\ref{section:mobility} were tested using the Android navigation app and the quantum web service, including the changes implemented after the initial testing phase in Wolfsburg. The significantly different conditions in Lisbon allowed us to further tune our navigation system, explained below.

{\bf Street exclusions.} The road network of Lisbon is significantly different from that of Wolfsburg. Apart from the major roads and highways in Lisbon, many of the local streets are narrow and one-way, making them not well suited for public transit. Additionally, many roads have steep inclines, which are difficult for buses to climb. Neither of these two conditions were present in Wolfsburg, which lead to multiple unacceptable scenarios when testing the buses in Lisbon. More than once, routes were suggested that utilized these small streets through which the buses could not fit, causing them to either turn back (adding delay to the travel time), or even worse, forcing them to stop completely. For example, between Saldanha roundabout and Alameda station there is a network of highly connected one-way streets. Suggested routes using these roads could be useful for cars, but were completely undesirable for the buses in the Quantum Shuttle fleet. Similarly, a network of narrow one-way streets exists near the Web Summit conference center, which also needed to be avoided. To accomplish this, whenever routes were suggested that utilized roads in these networks, the GPS locations were recorded, added to a list stored on the QWS server. This list (in the form of bounding boxes of areas to avoid on a map) was submitted as part of the HERE routing API request, which returned only routes that avoided those areas. This list of forbidden areas was actively updated throughout the successive tests in Lisbon based on feedback from the bus drivers, resulting in only valid routes being generated by the end of the final testing phase. This list of excluded regions was then used for the live run during the Web Summit. 

{\bf Time filtering.} The number of candidate routes that can be requested from the HERE routing API per vehicle is an unconstrained parameter client-side. By default, 3 candidate routes per vehicle were requested, to keep consistent with the work in~\cite{tfopaper}. However, it was observed that in some cases one or more of the routes suggested were significantly slower than the fastest route suggested, resulting in extremely slow routes sometimes being suggested. The reason for these slow routes being selected was due to the way the cost functions are formulated in the optimization problem. The goal is to reduce the amount of congestion caused by the vehicles, defined by the number of streets/GPS points shared between candidate routes across all vehicles. The slower routes suggested by the HERE routing API were often significantly longer than the fastest suggested route, and thus had lower overlap with the faster routes, causing some of the vehicles to be assigned to the slower routes. To circumvent this, a time filter was implemented to assure only reasonably fast routes were considered as valid candidates. After testing various values for the time filter, a value of 2~minutes provided the best trade-off between the number of routes selected and the routes' relative expected travel times. By allowing the slowest suggested route to be \emph{at most} 2~minutes slower than the fastest route suggested, we were able to maintain~3~valid candidates per vehicle for the majority of the trips. 

\section{Web Summit 2019: Live run}
\label{section:websummit2019}
The Quantum Shuttle service was active from November 4-7, 2019, and was operational for public use from November 5, 8:00 in the morning Lisbon time, to 18:00 in the evening on November 7, 2019. A total of 185 trips were recorded during the 4 day period with a total fleet size of 9 buses. However, a small number of trips were erroneously recorded, due to manual driver cancellation or restart of the trip. Such trips were identified in two ways: either a small number of vehicle locations recorded in the database (fewer than 5), or one of the origin/destination points being far from the expected location (more than 1 km). Of the 185 trips, 162 ($87.6\%$) were valid trips corresponding to the expected ``Quantum Shuttle'' service. The exact counts per day and line of the service are presented in Table~\ref{table:quantumshuttle}. As per Section~\ref{section:mobility} and shown in Figure~\ref{figure:shuttleroutes}, two of the lines operated from the city center of Lisbon to the Web Summit: Alameda station to Web Summit (blue line) and Saldanha roundabout to the Web Summit (red line\footnote{Since the quantum navigation of the green and red lines have identical origin and destination, we combine them and refer to them together as the red line.}). The third line (black) ran from the Web Summit to Alameda station.

\begin{table}[h]
    \centering
    \begin{tabular}{|c|c|c|c|}
        \hline
          & Black line & Blue line & Red line \\  \hline
         Conference total & 53 & 56 & 53 \\ \hline
         8:00-12:00 & 17 & 21 & 47 \\ \hline
         12:00-18:00 & 36 & 35 & 6 \\ \hline
    \end{tabular}
    \caption{The 162 trips taken by Carris buses operating the Quantum Shuttle, separated by time of day and line. }
    \label{table:quantumshuttle}
\end{table}

All three lines had roughly equal number of trips throughout the conference. The route frequency matches the expected demand in Lisbon-- the red line from Saldanha was the popular choice during the morning, whereas the blue line from Alameda was used more in the afternoon. Likewise, the only line from the Web Summit back to Lisbon city center (black line) had double the trip frequency in the afternoon compared to the morning, again matching the demand of conference attendees returning to their accommodations after the conference ends. 

The duration of each trip was recorded together with the location history of each vehicle in the fleet throughout the conference. The corresponding average trip times are shown in Table~\ref{table:routetimes}. The trip times are recorded from the moment a driver presses the start button, until either the trip is manually ended or the bus is within 50 meters of its destination.  

\begin{table}[h]
    \centering
    \begin{tabular}{|c|c|c|c|}
        \hline
          & Black line & Blue line & Red line \\  \hline
         Conference total & 23 min 41 s & 23 min 18 s & 26 min 34 s \\ \hline
         8:00-12:00 & 25 min 36 s & 22 min 43 s & 27 min 36 s \\ \hline
         12:00-18:00 & 22 min 46 s & 23 min 38 s & 18 min 19 s \\ \hline
    \end{tabular}
    \caption{Average trip times for the Quantum Shuttle, separated by time of day and line.}
    \label{table:routetimes}
\end{table}

One of the key design goals in our traffic navigation system was making sure it could operate continually  without manual intervention. As a consequence, there was significant variation in the complexity of the optimization problems being solved throughout the conference, depending on the number of active vehicles in the fleet. A total of 1275 optimization problems were solved by the QWS for the 162 trips, with an average response time of 4.69 seconds. Of those, 728 problems ($57.1\%$) involved more than one route per vehicle in the system, with an average response time of 6.78 seconds. We consider the optimization problems for which there is more than one route per vehicle as the ``harder'' version of the traffic flow problem, since otherwise the route selection is trivial. It is important to note that the ability to navigate the Quantum Shuttle buses strongly depended on creating and solving the optimization problems in a timely manner. Since we cannot anticipate in advance whether the vehicles have one or more possible routes (depending on the traffic conditions), our system needed to operate uninterrupted in all cases. Additionally, while there was a fallback mechanism in place as explained in Section~\ref{section:hexad}, $100\%$ of the calls to the D-Wave QPU completed successfully, thus maintaining our automated navigation system's integrity for the duration of the conference, regardless of the complexity of the problem being solved\footnote{The largest QUBO that was solved consisted of 12 variables, with 5 buses being navigated concurrently. This occurred on November 5, 9:11 Lisbon time, which was the busiest period during the conference.}.  

\begin{figure}[h]
  \centering
  \includegraphics[width=0.85\linewidth]{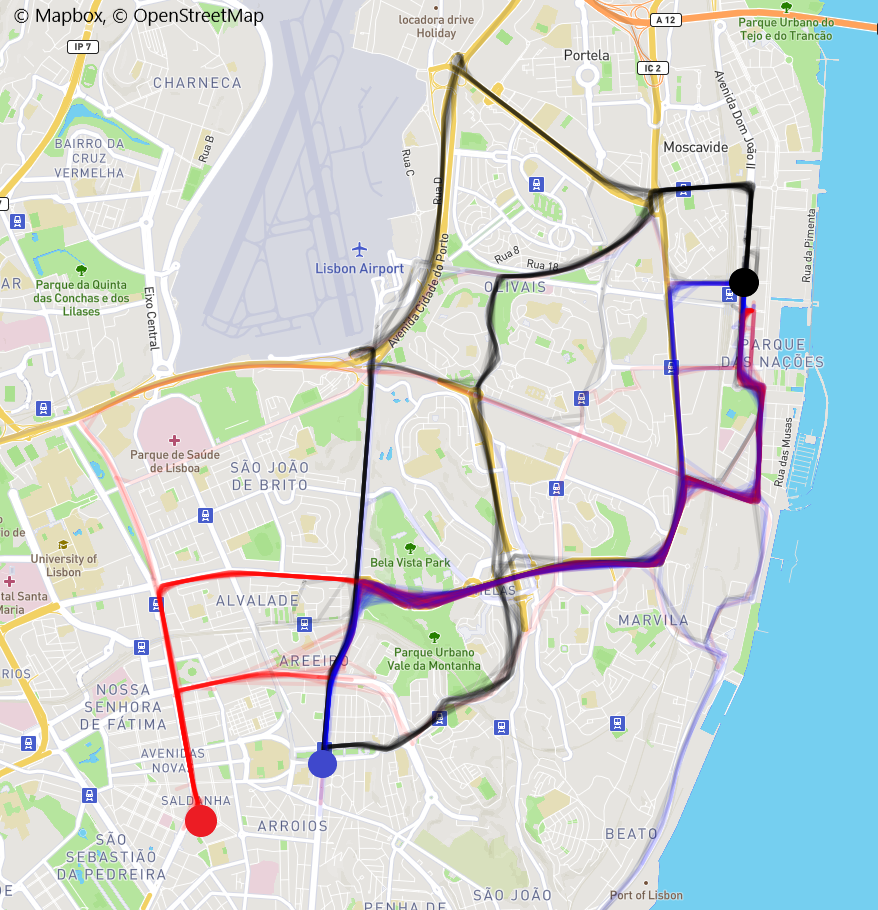}
  \caption{Distribution of all recorded Quantum Shuttle trips. The trips are color-coded based on the line they correspond to. The circles represent the origins and destinations of the respective lines~\cite{mapbox}. \label{figure:websummitroutesdata}}
  \Description{Heatmap of all data from Quantum Shuttle.}
\end{figure}

The distribution of all the Quantum Shuttle trip data is shown on a map of Lisbon in~Figure~\ref{figure:websummitroutesdata}. The trips are colored based on the lines to which they belong, as described above. The red, blue, and black circles correspond the the origins of their respective lines. The black circle is the Web Summit conference location, and is therefore also the destination for the red and blue lines. It is important to note that none of the three lines used the same route for all trips throughout the Web Summit, showing that our QWS navigation system provided flexible traffic-aware routing. The three highways that connect between the city center and the conference center were used extensively (although not exclusively), and at different times by the different lines. That these highways were prominent in the route selections is attributed to two design choices: time filtering and excluded streets. Highways are typically the fastest method of driving medium- and long-range distances, making them likely candidates for selection. Furthermore, the regions that were excluded from the route selection as per~Section~\ref{section:finaltests} removed fast route suggestions that avoided highways. It is reasonable to assume that in the case of navigating cars, as opposed to buses, Figure~\ref{figure:websummitroutesdata} would show increased distribution over the smaller city streets as well.

To quantify the customization of the routes used by the vehicles, we measure the dissimilarity between them. Specifically, we measure the overlap between the location histories of vehicles being navigated concurrently by our system. The overlap is defined as the fraction of GPS points in a vehicle's location history that coincided with another vehicle's route (that was being navigated at the same time), within a distance of 50 meters\footnote{This definition differs from the one in Section~\ref{section:tfo}, since the recorded location histories were irregular, as opposed to the points used to construct the TFO problem. The location histories were interpolated using the HERE API for consistency.}. The overlap metric is therefore defined to lie between [0, 1].  Distance $(d)$ between GPS points was calculated using the haversine formula:
\begin{equation}
    d = 2r\arcsin\left( \sqrt{\sin^2 \left( \frac{\phi_2 - \phi_1}{2}\right) + \cos{\phi_1}\cos{\phi_2}\sin^2 \left( \frac{\lambda_2 - \lambda_1}{2}    \right)}\right),
\end{equation}
where $\phi_1, \phi_2$ are latitudes, $\lambda_1, \lambda_2$ are longitudes, and $r = 6,371,009$ meters is the Earth's radius. 

\begin{figure}[h!]
  \includegraphics[width=\linewidth]{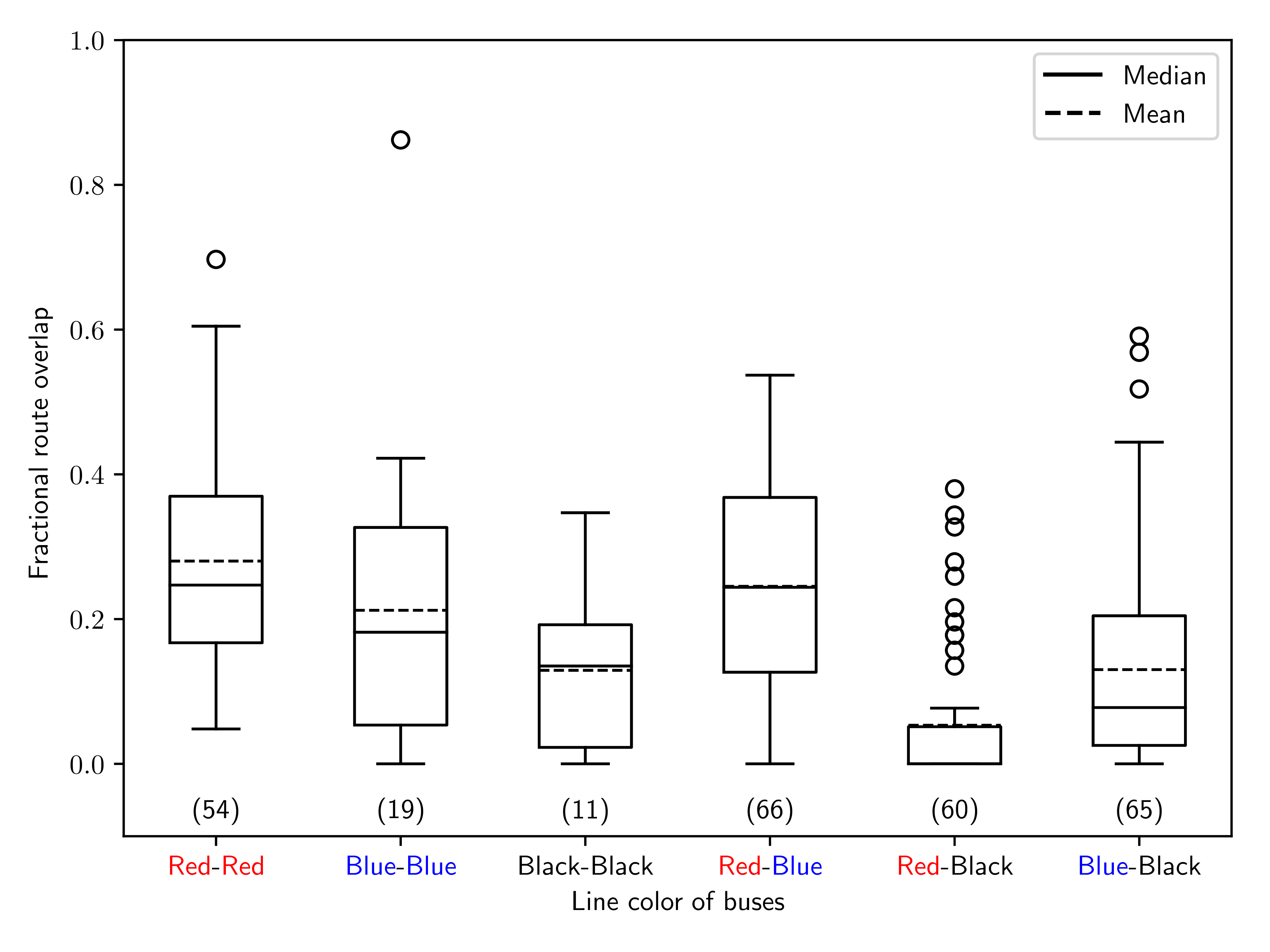}
  \caption{Box plot showing the fractional overlap of routes being navigated simultaneously in the Quantum Shuttle fleet. Boxes are grouped by the line colors. The number of observations are in parentheses below the boxes. \label{figure:overlap}}
  
\end{figure}

Because our system re-optimized the distribution of routes at most every 120 seconds, each vehicle's recorded trip is a sum of successive routes suggested by the optimization. It is important to note that the suggested routes, as obtained via the HERE API, are traffic-aware, and therefore already circumvent the existing traffic congestion in the city. Thus, by minimizing the overlap between the Quantum Shuttle vehicles, we can minimize the additional traffic congestion caused by our fleet. In Figure~\ref{figure:overlap} we show the overlap between pairs of buses, grouped by the lines the buses are following. There are six possible ways to compare the buses this way: three within the same line (red-red, blue-blue, black-black), and three between the lines (red-blue, red-black, and blue-black).

To contextualize the results in Figure~\ref{figure:overlap}, we also calculate the overlap between the most direct (i.e., fastest) routes suggested by the HERE API for each line without the traffic-aware component. This therefore simulates our navigation system in an ``offline'' mode-- public buses typically have pre-defined routes that are not deviated from even in the presence of traffic congestion. Using these static routes, we obtain the following overlaps: 0.70 for red-blue, 0.01 for red-black, and 0.16 for blue-black (intra-line overlaps are trivially 1, since the same routes would be used for every vehicle in that line). The red-black and blue-black overlaps are similar in offline and online mode, however the expected time for the offline mode would be higher due to the lack of traffic-aware routing. For the red-blue overlap (as well as the intra-route overlaps), our online method significantly reduces the overlap between the routes. It is also worth noting that in all categories in Figure~\ref{figure:overlap}, the mean and median overlaps were below 0.5, meaning that the majority of every route was different from every other vehicle in the fleet. By using the QWS we were therefore able to both circumvent existing congestion as well as avoid creating new congestion.

\section{Conclusions}
\label{section:conclusions}
In this work we presented our implementation of a live quantum computing-based navigation system for the Web Summit conference in Lisbon, Portugal. To our knowledge this is the first publicly accessible application that relied on live access to a quantum processor. Evidenced by the results presented (overlaps of routes and trip times) we were able to circumvent existing traffic in the city network while also minimizing the impact of our fleet on the city's traffic. Our main objectives, building a custom mobility service for a major event and using a quantum processor, were both achieved. 

The feedback from the drivers of the Quantum Shuttle fleet was overall mixed. One common criticism was that the front-end Android application was not optimized for in-route use, citing that it was difficult sometimes to start/end trips. Another comment was that it was difficult for some drivers to trust a remote system for optimized routing, as opposed to the ``offline'' approach of driving the same route for every trip. However, a large portion of drivers also commented that they were surprised at how effective the customized routing often was, despite the difficulties. Many of the drivers expressed their willingness to participate in an expanded trial of such a navigation system in the future. 

From a technological perspective, the Quantum Web Service's modular implementation allowed us to communicate with a live quantum processor in a timely fashion, making it suitable for our traffic optimization use-case. The test runs in Wolfsburg and Lisbon were particularly instrumental, allowing us to fine-tune the connections between the components of the QWS given the constraints of the application. Due to this, the final implementation of the Quantum Web Service can handle live optimization of other, non-traffic related processes provided modifications to the QUBO construction and live data components. This is of particular relevance to us, since many production processes have similar constraints to the those present in the Quantum Shuttle, making our work highly adaptable to other scenarios. In the future, we will use the same paradigm to test our Quantum Web Service for these different use-cases. 

There are still improvements that could be made to the QWS. Response times could be reduced by implementing a custom hybrid algorithm that can run in an asynchronously module as part of the QWS, as was proposed in Section~\ref{section:qpu}. Additionally, we could add a consolidated visualization platform to easily observe the operation of our system as a whole. These improvements and modifications, building on the existing QWS infrastructure, will be the main focus of our future work, and will allow us to truly start using quantum technologies for live production applications.



\begin{acks}
This work was funded through the Volkswagen Innovation Fund. We would like to thank Kelsey Hamer, Andy Mason, and Jeremy Hilton of D-Wave Systems, who offered continual support throughout the development of this project. We also thank Carris, C\^{a}mara Municipal de Lisboa, and PTV for their instrumental help in realizing this project. 
\end{acks}

\bibliographystyle{ACM-Reference-Format}
\bibliography{quantumshuttle}



\end{document}